\documentstyle[preprint,floats,aps,psfig]{revtex}
\tightenlines
\begin{document}

\draft

\title{ Form factors and photoproduction amplitudes}
\author{R. M. Davidson\thanks{davidr@rpi.edu} }
\address{Department of Physics, Applied Science and Astronomy \\
         Rensselaer Polytechnic Institute, Troy, New York 12180-3590}
\author{ Ron Workman\thanks{rworkman@gwu.edu} }
\address{Center for Nuclear Studies and Department of Physics \\
        The George Washington University Washington, DC 20052}

\draft
\date{\today}
\maketitle

\begin{abstract}

We examine the use of phenomenological
form factors in tree level amplitudes
for meson photoproduction. Two common recipes are shown
to be fundamentally incorrect. An alternate form consistent
with gauge invariance and crossing symmetry is proposed.

\end{abstract}
\vspace*{0.5in}

\pacs{PACS numbers: 25.20.Lj, 13.60.Le, 11.40.-q, 11.80.Cr}

\section{INTRODUCTION}
\label{Sec:Intro}

Studies of electromagnetic interactions with extended particles,
such as mesons and baryons, often involve the use of (off-shell) form factors
to account for internal structure not explicit in models, or to
regularize quantities which would otherwise be divergent. The general
structures of the $\pi$NN \cite{kazes} and $\gamma$NN \cite{binc}
three-point vertices
allowed based on the symmetries of the strong and electromagnetic interactions
have been known for decades. Of particular importance for electromagnetic
interactions is gauge invariance as expressed by the Ward-Takahashi
identities \cite{wt}. There is a vast literature on the subject of form
factors in meson production ranging from attempts \cite{west,gross} to
satisfy the half-off-shell Ward-Takahashi identities while at the same time
using as much on-shell information as possible, to explicit model
calculations \cite{naus,bos} which test the various prescriptions. As a rule,
these model calculations demonstrate that there are kinematical regions
where the prescriptions can give drastically different results than the
exact model calculations.

While the most consistent way to approach this problem is through
the use of field-theoretic models, it is not clear which field-theory
approach to use in the resonance region. Although chiral perturbation
theory can be used near threshold \cite{bern1}, it is expected to
converge slowly, or not at all, at higher energies. Thus,
many recent fits to meson
photoproduction data have been less ambitious, and have assumed
that a phenomenological approach would be adequate for the 
extraction of resonance contributions.
These fits have generally used tree-level diagrams which require
a high-energy cutoff. While the functional form of these
form factors has usually been chosen on the basis of convenience and
simplicity, the resulting amplitudes should obey the constraints 
imposed by gauge invariance, crossing and unitarity. Most efforts have
been directed toward the maintenance of gauge invariance. Crossing
and unitarity are often ignored in modifications applied to tree-level
diagrams, whereas unitarity is generally built into dynamical models.

In the following, we will concentrate on the modification of tree-level
diagrams. The effect of a minimal-substitution prescription, applied to
the most general (pion-nucleon) vertex, was discussed in detail by
Ohta\cite{ohta} and extended by others\cite{haberzettl,workman,wang}.
This work showed that a
particular invariant amplitude ($A_2$) would be unaltered by a wide
range of form factors and meson-nucleon vertices satisfying the
Ward-Takahashi\cite{wt} identity. The possibility of having a
form-factor modification of $A_2$, accompanied by a gauge-invariance
restoring contact term was subsequently considered in 
Ref.\cite{haberzettl}. The recipe of Ref.\cite{haberzettl}
is simple to apply and has been widely adopted for use in single-
and multi-channel fits to meson-photoproduction data \cite{feuster}.

We find a flaw in the arguments used to derive this modification of
the amplitude $A_2$. It is, however, straightforward to find a
modification of $A_2$ which {\it does} satisfy the constraints of
gauge invariance and crossing symmetry. We then apply this technique
to $\eta^{\prime}$ photoproduction to overcome the shortcomings of
Refs.~\cite{nimai,muk}, namely the violation of crossing symmetry.
We also show how the use of our prescription can alter the
conclusions of a phenomenological analysis.

\section{REVIEW OF THE $A_2$ PROBLEM}
\label{Sec:Form}
We begin by recalling the basic results given in Ref.\cite{workman}. There
the discussion was simplified by taking the simplest (pseudoscalar) 
$\pi NN$ vertex in the specific process $\gamma p \to n\pi^+$. The contribution
from the Born diagrams corresponding to $s-$, $t-$, and $u-$channel
exchanges is
\begin{eqnarray}
\epsilon \cdot M_{fi} & = & \sqrt{2}ge \bar u_n \gamma_5
{ { (p_1+k)\cdot \gamma + m }\over {s - m^2} } \left[
\epsilon \cdot \gamma - { \kappa_p \over {2m} } \epsilon \cdot \gamma
k \cdot \gamma \right] u_p \\
 & + & 2\sqrt{2}ge \bar u_n { {q\cdot \epsilon}\over{t - \mu^2} } \gamma_5 u_p 
\nonumber \\
 & - & \sqrt{2}ge \bar u_n {\kappa_n \over {2m}}
 \epsilon \cdot \gamma k\cdot \gamma
{{(p_2-k)\cdot \gamma + m }\over {u-m^2} } \gamma_5 u_p , \nonumber
\end{eqnarray}
where $k$ and $q$ represent the photon and pion
four-momenta, and $p_1$ and $p_2$ are the respective proton
and neutron four-momenta. The quantities $m$ and $\mu$ are the nucleon
and pion masses, $\epsilon$ is the photon polarization vector, $g$ is the
pseudoscalar $\pi^0 pp$ coupling constant, and $\kappa_p$ and $\kappa_n$ are
the proton and neutron anomalous magnetic moments. Note that in
Refs.~\cite{haberzettl,workman}, $g$ denoted the pseudoscalar $\pi^+ np$
coupling constant.

If a strong form factor is inserted at the $\pi NN$ vertex of each Born
term, gauge invariance is clearly lost. The following term violates gauge
invariance:
\begin{equation}
\epsilon \cdot \bar M = \sqrt{2}ge\bar u_n \gamma_5 
\left[ { {2p_1 \cdot \epsilon} \over 
{s-m^2} } F(s , m^2 , \mu^2) + {{2 q\cdot \epsilon } \over {t-\mu^2} }
F(m^2 , m^2 ,t) \right] u_p \; ,
\end{equation}
where $F(s,u,t)$ is a general form factor. From Ohta's\cite{ohta} relations,
the term required to restore gauge invariance is
\begin{equation}
\sqrt{2}ge\bar u_n \gamma_5 \left[ { {2p_1 \cdot \epsilon} \over
{s-m^2} } [ F(m^2,m^2,\mu^2) - F(s, m^2 , \mu^2 )]  
 + {{2 q\cdot \epsilon } \over {t-\mu^2} }
[ F(m^2,m^2 , \mu^2) - F(m^2 , m^2 ,t) ] \right] u_p, 
\end{equation}
with $F(m^2,m^2,\mu^2)=1$. This precisely cancels the form factor 
effect on the terms in Eq.~2. Writing this in terms of invariant
amplitudes
\begin{equation}
\epsilon \cdot M_{fi} = \bar u_n \sum_{j=1}^4 A_j M_j u_p,
\end{equation}
with the explicitly gauge invariant representation
\begin{eqnarray}\label{invamp}
M_1 & = & -\gamma_5 \epsilon \cdot \gamma k\cdot \gamma ,\\
M_2 & = & 2\gamma_5 ( \epsilon \cdot p_1 k\cdot p_2 
                       - \epsilon \cdot p_2 k\cdot p_1 ) \nonumber , \\
M_3 & = & \gamma_5 ( \epsilon \cdot \gamma k\cdot p_1 
                       - \epsilon \cdot p_1 k\cdot \gamma ) \nonumber , \\
M_4 & = & \gamma_5 ( \epsilon \cdot \gamma k\cdot p_2
                       - \epsilon \cdot p_2 k\cdot \gamma ) \nonumber ,
\end{eqnarray}
noting that such a representation is only possible for a gauge invariant 
amplitude, we have
\begin{eqnarray}\label{as}
A_1 & = & { {\sqrt{2}geF(s,m^2 , \mu^2)}\over {s-m^2}} (1+\kappa_p) +
{{\sqrt{2}geF(m^2 ,u, \mu^2)}\over {u-m^2}} \kappa_n, \\ 
A_2 & = & {{2\sqrt{2}ge}\over {(s-m^2)(t-\mu^2)} }, \nonumber \\
A_3 & = & {{\sqrt{2}ge}F(s, m^2 , \mu^2 )\over {s-m^2}} {\kappa_p \over m},
\nonumber \\
A_4 & = & {{\sqrt{2}ge}F(m^2 ,u, \mu^2)\over {u-m^2}} {\kappa_n \over m}
\nonumber .
\end{eqnarray}
Thus, one sees that the $A_2$ amplitude is unaltered by form factors,
within
Ohta's prescription for retaining gauge invariance. The result holds for
a more general pion-nucleon vertex and any charge channel. Of course, this
result does not prevent the addition of terms which are individually
gauge invariant. 

Since form factors were added to provide a cutoff for the Born term
contributions,
this result is somewhat disappointing, as the $A_2$
contribution is not damped. In addition, explicit one-loop meson
calculations \cite{bos,bern1} show that the $A_2$ amplitude is modified.
In Ref.\cite{haberzettl}, a form factor for $A_2$
was inserted and the addition of a contact interaction 
was proposed to restore gauge invariance. We give this recipe below, show why
it has a problem, and then propose an alternate form. The result is then
generalized using crossing symmetry.

\section{THE HABERZETTL FORM FACTOR}

In Ref.~\cite{haberzettl} a gauge invariant term,
$\hat F A_2$, was added and subtracted
from a set of Born terms with form factors at the strong vertices. This 
resulted in the replacement $A_2 \to \hat F A_2$ above, with a remaining
gauge-invariance-violating term
\begin{eqnarray}
\epsilon \cdot M_{\rm viol} & = & \sqrt{2}
ge \bar u_n \gamma_5 \left[ {{2p_1 \cdot \epsilon}\over {s-m^2}}
(F(s,m^2,\mu^2) - \hat F ) \right. \\
 & + & \left. {{2q\cdot \epsilon}\over {t-\mu^2}} 
(F(m^2,m^2,t) - \hat F ) \right] u_p \; , \nonumber
\end{eqnarray}
which was to be cancelled by a {\it contact} interaction.

By requiring this additional term to be free of poles, i.e., a contact term,
we can constrain 
the functional form of $\hat F$. Denoting $F(s,\mu^2,m^2) \equiv F_1 (s)$,
with similar abbreviations for $F_2(u)$ and $F_3(t)$, subject to
the constraints
\begin{equation}
F_1 (m^2) =F_2 (m^2 ) =F_3 (\mu^2) = 1 \; ,
\end{equation}
the general form for $\hat F$ used in Ref.\cite{haberzettl} was
\begin{equation}
\hat F = a_1 F_1(s) + a_2 F_2(u) + a_3 F_3(t) \; ,
\end{equation}
subject to $a_1 + a_2 + a_3 = 1$. This form has been used in many recent
analyses of photonic and multi-channel reactions off the
nucleon \cite{feuster}. However,
it is straightforward to demonstrate that this recipe
results in a term with pole contributions, and is therefore incorrect
because it cannot be cancelled by a contact term.
Taking $a_2 = 1$, and inserting
$\hat F$ into the relation for $M_{\rm viol}$, one can easily see the
problem. Choosing, for example, the point $s=m^2$ does not fix the value 
of $u$, which is only constrained by the condition $t+u = \mu^2 + m^2$. As a 
consequence, when $s = m^2$ it is not necessarily the case that $F_2 (u)$ =1.
As a result, the
pole terms are not cancelled, except for zero photon energy, as noted by the
authors. To ensure cancellation of the poles, the needed constraint is, in this
case,
\begin{equation}
\hat F (s=m^2 , t) = \hat F (s, t=\mu^2 ) =1 \; ,
\end{equation}
which does not, however, imply that $\hat F (s,t)$ = 1 for all $s$ and $t$.
For example, the
following choice\footnote{It should be noted that this choice is not unique.}
\begin{equation}\label{fhat}
\hat F = F_1 (s) + F_3 (t) - F_1 (s)F_3 (t)
\end{equation}
satisfies the above constraint. Note that this method can be easily
adapted to charged kaon production.

So far, we have only considered constraints provided by gauge invariance
and the pole structure of the amplitude. However, crossing symmetry is
also important.
As the $p \pi^-$ channel is closely related to the $n\pi^+$ channel
via crossing, let us therefore consider the isospin channel $T^{(-)}$
defined by\footnote{The isoscalar channel, $T^{(0)}$, is discussed below.}
\begin{equation}
T^{(-)} = { 1 \over 2\sqrt{2}} ( T^{n\pi^+} - T^{p\pi^-}) \; .
\end{equation}
Inserting strong form factors, the matrix element for the
electric Born terms is
\begin{eqnarray}
\epsilon \cdot M &=& {eg \over 2} F_1 (s) \bar{u}_2 \gamma_5
{ ( p_1 + k )\cdot \gamma +m \over s-m^2 }
\gamma \cdot \epsilon u_1 \nonumber \\
&-& {eg \over 2}F_2 (u) \bar{u}_2 \gamma \cdot \epsilon { ( p_2 -
k )\cdot \gamma +m \over u-m^2 } \gamma_5 u_1 \nonumber \\
&+& 2eg F_3 (t) \bar{u}_2 \gamma_5 { q \cdot \epsilon \over
t-\mu^2 } u_1 \; .
\end{eqnarray}
After some rearrangement, this may be written as
\begin{eqnarray}
\epsilon \cdot M &=& {eg \over 2} \bar{u}_2 M_1 \left[ { F_1 (s) \over
s-m^2 } - { F_2 (u) \over u-m^2 } \right] u_1 \nonumber \\
&+& eg \hat F (s,u,t) \bar{u}_2 {M_2 \over (t-\mu^2)}
\left[ { 1 \over s-m^2} - { 1 \over u-m^2} \right] u_1 \nonumber \\
&+& \epsilon \cdot \Delta M \; ,
\end{eqnarray}
where $M_{1,2}$ are the gauge invariant operators 
defined in Eq.~\ref{invamp} and
\begin{eqnarray}
\epsilon \cdot \Delta M &=& -{eg \over 2} \bar{u}_2 \gamma_5 \left[
{ 2p_1 \cdot \epsilon \over s-m^2 } (\hat F (s,u,t) -F_1 (s) ) \right.
\nonumber \\
 & + & \left. { 4q \cdot \epsilon \over t-\mu^2 } (\hat F (s,u,t) -F_3 (t) )
- { 2p_2 \cdot \epsilon \over u-m^2 } (\hat F (s,u,t) -F_2 (u) )
\right] u_1 \; .
\end{eqnarray}
As $A_{1,2}^{(-)}$ are odd under crossing, i.e.,
\begin{equation}
A_{1,2}^{(-)} (s,u,t) = - A_{1,2}^{(-)} (u,s,t) \; ,
\end{equation}
we see that $\hat F (s,u,t)$ = $\hat F (u,s,t)$ and $F_2 (u)$ =
$F_1 (u)$, that is, $F_2 (u)$ is obtained from $F_1 (s)$ by the
replacement $s$ $\rightarrow$ $u$. In order to cancel the poles in
$\epsilon \cdot \Delta M$, so this term may be cancelled by a contact term,
we need
\begin{equation}
\hat F (m^2 , u,t) = \hat F (s , m^2 ,t) = \hat F (s , u,\mu^2) = 1 \; .
\end{equation}
Although not unique, one crossing symmetric form for $\hat F$ that
satisfies the above constraints is
\begin{eqnarray}
\hat F (s,u,t) &=& F_1 (s) +F_1 (u) +F_3(t) \nonumber \\
&-& F_1 (s) F_1 (u) - F_1 (s) F_3 (t) -F_1 (u) F_3 (t)
+ F_1 (s) F_1 (u) F_3 (t) \; .
\end{eqnarray}

\section{PHENOMENOLOGY}

The use of different form factor schemes can have a significant effect on
phenomenological fits to data. This is particularly clear in kaon
photoproduction where
the rapid increase of a point-like Born contribution to this quantity
is well known. The point-like Born term contribution
to the total cross section (solid) is compared to the data \cite{said}
in Fig.~1 using a $K\Lambda N$ coupling of 7.5 \cite{klam}.
An attempt to cure this, by modifying individual multipoles,
was described in Ref.\cite{tkb}. Using
the minimal substitution prescription of Ohta\cite{ohta} (dotted),
which damps all
but the $A_2$ amplitude, one still obtains a Born contribution which grows too
rapidly. Here, we use
\begin{equation}\label{ff}
F_1 (s) = [ 1 + (s-m^2 )^2 / \Lambda^4 ]^{-1} \; ,
\end{equation}
with a cutoff mass $\Lambda$ of 1 GeV. Finally, we show in Fig.~1
the result arising from our modification of the Haberzettl form (dashed).
In addition to $F_1 (s)$ given above, we also need $\hat{F}$ and $F_3 (t)$.
We take $\hat{F}$ to be of the form of Eq.~(\ref{fhat}) with
\begin{equation}
F_3 (t) = [ 1 + (t-\mu^2 )^2 / \Lambda^4 ]^{-1} \; ,
\end{equation}
where $\mu$ is the kaon mass and we again take $\Lambda$ = 1 GeV.

Finally, we consider the fits to $\eta^{\prime}$
photoproduction performed in
Refs.\cite{nimai,muk}. In these studies, the Born terms were added and
compared to
possible resonance contributions. The use of a coupling, $g_{\eta^{\prime}}$,
within the range of values, $1\le g_{\eta^{\prime}} \le 6$, estimated in this
work\cite{nimai} results in an (undamped) Born contribution exceeding the
data \cite{data,ahhm,bonn} by a large factor, as is demonstrated by the solid
line in Fig.~2 (here, $g_{\eta^{\prime}}$ = 1.9). There is more than
one way to interpret this result. One possibility is that $g_{\eta^{\prime}}$
is much smaller than estimated in Ref.~\cite{nimai}. A different possibility
is that $g_{\eta^{\prime}}$ is in the range given above, but the
{\it effective} $g_{\eta^{\prime}}$ is much smaller than $g_{\eta^{\prime}}$
at the nucleon pole. In other words, as the proton and eta-prime are extended
objects, form factors enter at the three-point vertices, and as these
data are far from the nucleon pole, form factor effects can be significant.
Finally, there are almost certainly other reaction mechanisms such as
resonance exchanges, vector meson exchanges, and possibly contact terms.

In the original analysis of the $\eta^{\prime}$ data
\cite{nimai}, the Born terms were multiplied by an overall
form factor given by Eq.~(\ref{ff})
with $\Lambda$ = 1.1 GeV. In this model, the dominant contribution
to the model was from the {\it background} of the D13(2080) resonance, and
the Born terms played a minor role. The resulting total cross section from
the Born terms is shown by the dashed (nearly zero) line in Fig.~2. On the
other hand, taking $g_{\eta^{\prime}}$ = 3.5 and $\Lambda$ = 2 GeV,
it was shown
that one could qualitatively reproduce the total cross section with just
Born terms (dotted line) \cite{muk}. However,
the procedure of multiplying the Born terms by an overall form factor
violates crossing
symmetry and the amplitude has the wrong residue structure, and therefore is
theoretically unacceptable. Other inconsistencies of this procedure
have been discussed in Refs.~\cite{naus,bos}. Thus, it is not clear
what conclusions, if any, can be drawn from the dashed and dotted curves
in Fig.~2.

Using Ohta's
\cite{ohta}
method, one obtains a crossing symmetric amplitude, but as above the $A_2$
amplitude is not modified. Using our modification of the Haberzettl form,
the electric Born terms are
\begin{eqnarray}\label{etaamp}
A_1 &=& e g_{\eta^{\prime}} \left[ {F_1 (s) \over s-m^2 } +
{ F_1 (u) \over u-m^2 } \right] \; , \nonumber \\
A_2 &=& { 2eg_{\eta^{\prime}} \over (s-m^2)(u-m^2) } \hat{F}(s,u)\; ,
\end{eqnarray}
where $F_1 (m^2 )$ = 1.
In this case, the constraints on $\hat{F}(s,u)$ are
\begin{equation}
\hat{F}(s,m^2 ) = \hat{F}(m^2 ,u) = 1 \; ,
\end{equation}
from the condition to remove the poles, and $\hat{F}(s,u)$ = $\hat{F}(u,s)$
from crossing symmetry. Note that this form is also valid for $\pi^0$
and $\eta$ production, and the electric Born terms for isoscalar production
of pions, defined by
\begin{equation}
T^{(0)} = { 1 \over 2\sqrt{2}} ( T^{n\pi^+} + T^{p\pi^-}) \; ,
\end{equation}
can be obtained from Eq.~(\ref{etaamp}) using the replacement
$g_\eta^{\prime}$ $\rightarrow$ $g$/2.
With this form, the Born terms in $\eta^{\prime}$ production
can be sufficiently damped at high energies even for the largest estimated
value of $g_\eta^{\prime}$. For example, taking
\begin{equation}
\hat{F} = F_1 (s) + F_1 (u) -F_1 (s)F_1 (u) \; ,
\end{equation}
with $F_1 (s)$ given by Eq.~(\ref{ff}), and using the
upper limit for $g_{\eta^{\prime}}$ with a cutoff of 700 MeV,
the total cross section
from the Born terms levels off at about 0.2 $\mu$b at a photon energy of
2 GeV. Furthermore, and this is the essential point, it is {\it not possible}
to reproduce the total cross section shape here with only Born terms and
reasonable values for $\Lambda$ (500 to 2000 MeV), in contrast to the
non-crossing symmetry model of Ref.\cite{muk}. Thus, the use of the 
present, more restrictive, form factor prescription narrows the possible
interpretations of experimental data.

\section{CONNECTION WITH CONTACT TERMS}

It is known in field theory \cite{stef}
that off-shell form factors are not unique and
are closely related to contact terms. For different representations of the
fields, the form factors and contact terms differ, but the differences
cancel out in the calculation of a physical scattering amplitude. In the
case of dispersion relations \cite{kazes,binc},
the nonuniqueness of the off-shell form factors
is reflected in the a priori unknown number of needed subtractions \cite{dav}.
For the
phenomenological amplitudes presented here, it is also possible to shift
form factor effects to contact terms. To illustrate this, recall that
$F_1 (m^2)$ = 1 and we may write, for
example, for the $A_1$ term in Eq. (\ref{etaamp}),
\begin{equation}
A_1 = e g_{\eta^{\prime}} \left[ {1 \over s-m^2 } +
{ 1 \over u-m^2 } +
{( F_1 (s)-1) \over s-m^2 } +
{ (F_1 (u)-1) \over u-m^2 } 
\right] \; .
\end{equation}
The first two terms above are recognized as the undamped Born terms, while
the last two terms correspond to contact terms since the pole is cancelled
as $s$ or $u$ $\rightarrow$ $m^2$. As previously noted, the $A_2$ amplitude
is not modified in Ohta's method \cite{ohta}, but his method does not forbid
the inclusion of an additional gauge invariant contact term. Our modification
of Haberzettl's procedure is equivalent to adding the gauge invariant
contact term,
\begin{equation}
A_2 \bar{u}_2 M_2 u_1 (\hat{F} -1) \; ,
\end{equation}
to the amplitude of Ref.~\cite{workman} (Eq. (\ref{as}) of this work).
Here, $M_2$ is the gauge invariant operator defined in
Eq. (\ref{invamp}) and $A_2$ is given by the second
equation of Eqs. (\ref{as}).

Finally, let us note that most authors assume that the form factors are
real, and the phenomenological
results presented here are also obtained using real form factors.
In general, however, arguments based on unitarity \cite{kazes,binc,dav}
show that the off-shell form factors must be complex above the one-pion
threshold. On the other hand, there is some justification for using
real form factors if one adopts a K-matrix approach to the scattering amplitude
\cite{kon}.

\section{CONCLUSIONS}

We have found a flaw in the gauging procedure of Ref.~\cite{haberzettl} and
have noted that the method used in Refs.~\cite{nimai,muk}
is theoretically
unacceptable as it violates crossing symmetry. We have determined
the general constraints $\hat{F}$ must satisfy for the gauge-violating
term to be a contact term, which then can be cancelled by a contact term.
This method was generalized to be consistent with crossing symmetry and
applied to $\eta^{\prime}$ photoproduction. This model is theoretically
superior to previous works \cite{nimai,muk} and has the important
phenomenological consequence that the total cross section cannot be
interpreted as arising solely from the Born terms, in contrast to
Ref.~\cite{muk}. Other phenomenological applications are being investigated.

\acknowledgments

We are grateful to Nimai C. Mukhopadhyay for many useful discussions
regarding eta-prime photoproduction and the need for a crossing symmetric
model. We thank Paul Stoler for a critical reading of the manuscript.
This work was supported in part by the U.~S. Department of 
Energy Grants DE--FG02--99ER41110 and DE--FG02--88ER40448.
R.W. gratefully 
acknowledges a contract from Jefferson Lab under which 
this work was done.  Jefferson Lab is operated by the 
Southeastern Universities Research Association under the
U.~S.~Department of Energy Contract DE--AC05--84ER40150.
R.M.D. thanks the Center for Nuclear Studies for supporting a visit
in which part of this work was completed.

\eject


\begin{figure}
\caption{The total cross section for $\gamma p$ $\rightarrow$
$K^+ \Lambda$ versus the photon lab-energy, $E_{\gamma}$. Plotted
are the unmodified Born terms (solid), Ohta's prescription
\protect\cite{ohta} (dotted), and our form-factor modification
(see text). Data from Ref.\protect\cite{said}.}
\caption{The total cross section for $\gamma p$ $\rightarrow$
$p \eta^{\prime}$ versus the photon lab-energy, $E_{\gamma}$.
The data are from SAPHIR \protect\cite{bonn} (diamonds), AHHM
\protect\cite{ahhm} (squares) and ABBHHM \protect\cite{data} (crosses).
The solid line is the total cross section obtained from the Born terms
with $g_{\eta^{\prime}}$ = 1.9 and without any form factors. The dashed line
is from the work of Ref.~\protect\cite{nimai} where the Born terms are
multiplied by an overall form factor depending only on $s$, thus violating
crossing symmetry. The dotted line is the best Born model fit to the
data using the non-crossing symmetric model \protect\cite{muk}.}
\end{figure}

\vfill
\eject


\begin{thebibliography}{99}

\bibitem{kazes} E. Kazes, Nuov. Cim. {\bf 13}, 1226 (1959).

\bibitem{binc} A. Bincer, Phys. Rev. {\bf 118}, 855 (1960).

\bibitem{wt}
J.C. Ward, 
Phys. Rev. {\bf 78}, 182 (1950);
Y. Takahashi, Nuovo Cimento {\bf 6}, 371 (1957).

\bibitem{west} F.A. Berends and G.B. West, Phys. Rev. {\bf 188}, 2538 (1969).

\bibitem{gross} F. Gross and D. Riska, Phys. Rev. C {\bf 36},
1928 (1987).

\bibitem{naus} H.W.L. Naus and J.H. Koch, Phys. Rev. C {\bf 39}, 1907
(1989).

\bibitem{bos} J.W. Bos, S. Scherer and J.H. Koch, Nucl. Phys. {\bf A547},
488 (1992).

\bibitem{bern1} V. Bernard, N. Kaiser, and U.-G. Meissner, Phys. Lett.
{\bf B282}, 448 (1992); {\it ibid}, {\bf B378}, 337 (1996).

\bibitem{ohta}
K. Ohta,
Phys. Rev. C {\bf 40}, 1335 (1989).

\bibitem{haberzettl}
H. Haberzettl, C. Bennhold, T. Mart, and T. Feuster,
Phys. Rev. C {\bf 58}, R40 (1998).

\bibitem{workman}
R.L. Workman, H.W.L. Naus, and S.J. Pollock,
Phys. Rev. C {\bf 45}, 2511 (1992).

\bibitem{wang}
S. Wang and M.K. Banerjee,
Phys. Rev. C {\bf 54}, 2883 (1996).

\bibitem{feuster} T. Feuster and U. Mosel, Phys. Rev. C {\bf 59}, 460 (1999);
T.~Mart and C.~Bennhold, Phys. Rev. C {\bf 61}, 012201(R) (1999);
T.~Mart, Phys. Rev. C {\bf 62}, 038201 (2000);
F.X. Lee, T.~Mart, C.~Bennhold, H.~Haberzettl, L.E.~Wright,
nucl-th/9907119.

\bibitem{nimai}J.-F. Zhang, N.C. Mukhopadhyay and M. Benmerrouche, Phys.
Rev. C {\bf 52}, 1134 (1995).

\bibitem{muk}N.C. Mukhopadhyay, J.-F. Zhang, R.M. Davidson and M. Benmerrouche,
Phys. Lett. {\bf B410}, 73 (1997).

\bibitem{said}
Data from the SAID database \hbox{(http://gwdac.phys.gwu.edu)}. This
source also contains representative fits from several different
groups.

\bibitem{klam} This value for the $K \Lambda N$ coupling is within the range of 
values found or used in: R.A. Adelseck, C. Bennhold, and L.E. Wright,
Phys. Rev. {\bf C} 32, 1681 (1985) and Ref. \protect\cite{haberzettl}. 

\bibitem{tkb}
H. Tanabe, M. Kohno and C. Bennhold,
Phys. Rev. C {\bf 39}, 741 (1989).

\bibitem{data} The ABBHHM collaboration, Phys. Rev.
175, 1669 (1968).

\bibitem{ahhm} The AHHM collaboration, Nucl. Phys. {\bf B108}, 45 (1976).

\bibitem{bonn} R. Pl{\"o}tzke {\it et al}., Phys. Lett.
{\bf B444}, 555 (1998).

\bibitem{stef} S. Scherer and H. Fearing, Phys. Rev. C {\bf 51}, 359 (1995).

\bibitem{dav}R.M. Davidson and G.I. Poulis,
Phys. Rev. D {\bf 54}, 2228 (1996).

\bibitem{kon} S. Kondratyuk and O. Scholten, Phys. Rev. C {\bf 59},
1070 (1999).

\end{thebibliography}
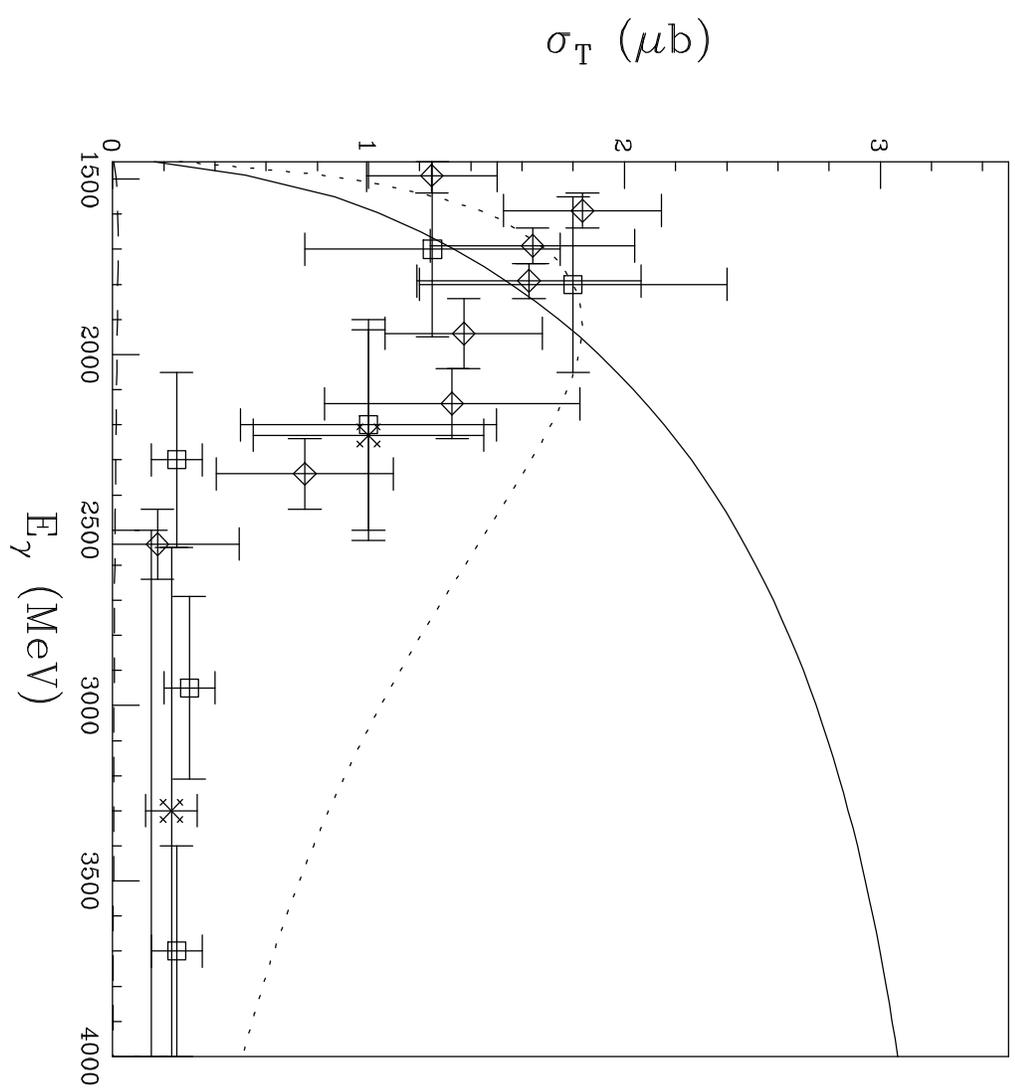
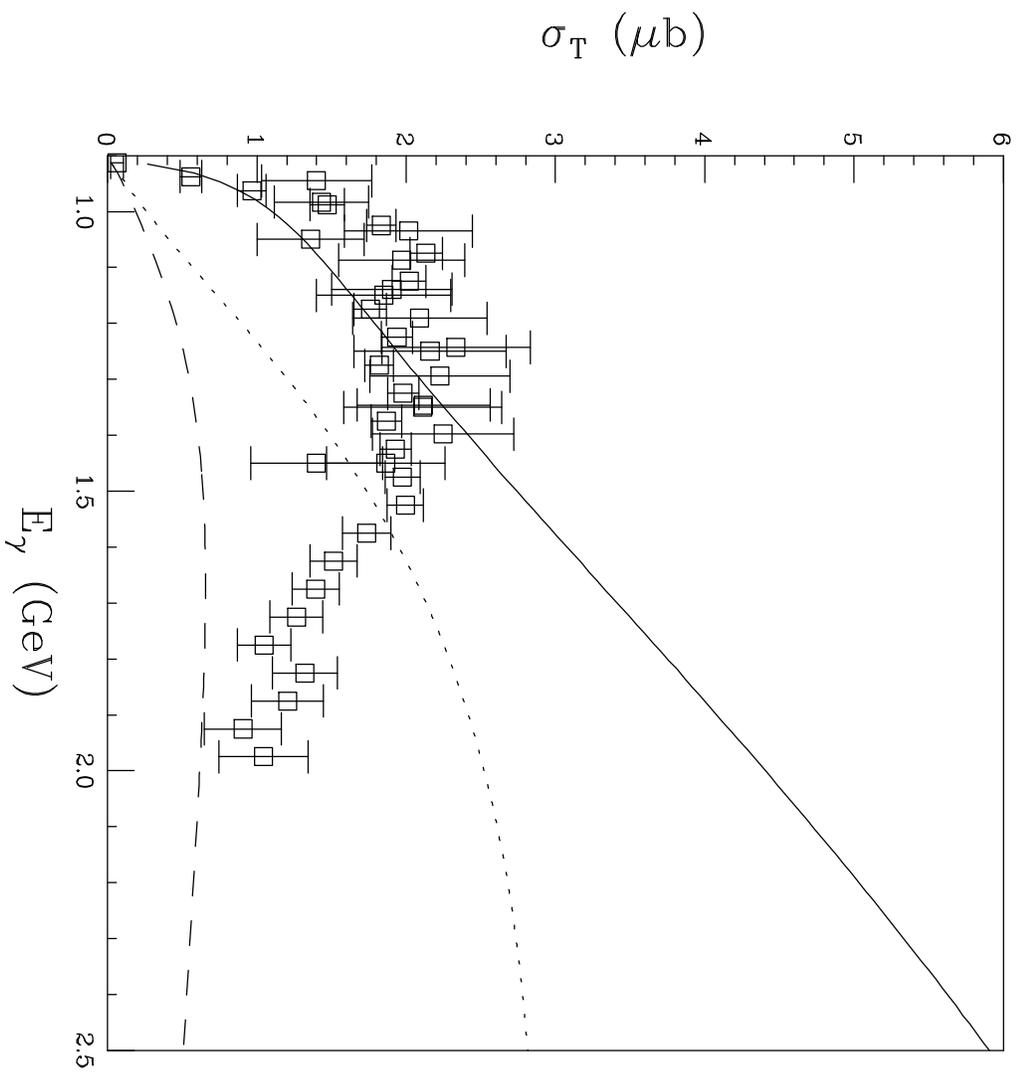
\end{document}